\documentstyle[proceedings,psfig]{crckapb}
\begin{opening}
\title{Hierarchical Systems in Open Clusters}
\author{R. de la Fuente Marcos}  
\institute{Universidad Complutense de Madrid, E-28040, Madrid, Spain 
           \\ fiast05@emducms1.sis.ucm.es}
\author{S. J. Aarseth}
\author{L. G. Kiseleva}
\author{P. P. Eggleton}  
\institute{Institute of Astronomy, \\ Madingley Rd., Cambridge CB3 0HA, UK}
\end{opening}

\begin{document}

\begin{abstract} 
In this paper we study the formation, evolution 
and disruption of hierarchical systems in open clusters.
With this purpose, $N$-body
simulations of star clusters containing an initial population of binaries
have been carried out using Aarseth's NBODY4 and NBODY5 codes. 

Stable triples may form from strong interactions of two binaries in which
the widest pair is disrupted.
The most frequent type of hierarchical systems found in the cluster models
are triples in which the outer star is single, but in some cases the outer
body is also a binary, giving a hierarchical quadruple.
The formation of hierarchical systems of even higher multiplicity is also
possible. 
Many triple systems are
non-coplanar and the presence of even a very distant and small
outer companion may affect the orbital parameters
of the inner binary, including a possible mechanism of
significant shrinkage if the binary experiences a weak tidal dissipation.

The main features of these systems are analyzed
in order to derive general properties which can be
checked by observations. 
The inner binaries have periods in the range 1 -- 10$^3$ days, although rich 
clusters may have even smaller periods following
common envelope evolution.
For triple systems, the outer body usually has a mass less than 1/3 
of the binary,
but is sometimes a collapsed object with even smaller mass.
The formation of exotic objects, such as blue stragglers
and white dwarf binaries, inside hierarchical triple systems is
particularly interesting. 
An efficient mechanism for generating 
such objects is the previous 
formation of a hierarchical system in which the inner binary may develop a very 
short period
during a common envelope phase, which finally results in
a stellar collision.
\end{abstract}

\section{Introduction}
In the past few years significant evidence for a large number of binary
and multiple stellar systems in galactic clusters has been obtained from 
observational surveys (Mathieu et al. 1990; Mermilliod et al. 1992;
Ghez et al. 1993; Mason et al. 1993; Mayer et al. 1994; Mermilliod et al. 1994;
Simon et al. 1995), both for zero age main sequence and 
pre-main-sequence stars. For multiple systems the frequency estimated 
for star forming regions is up to 35 \% and for the field is up to 20 \%.
However, the fraction of these systems detected in open clusters is 
smaller, although the improvement in observational techniques has
increased in the past few years. Currently, the majority of 
multiple systems discovered in open clusters are triples and quadruples.
These systems are usually highly hierarchical. Triple (or even higher
multiplicity) systems are found in the Pleiades (Mermilliod et al. 1992),
the Hyades (Griffin \& Gunn 1981, Griffin et al. 1985, Mason et al. 1993),
Praesepe (Mermilliod et al. 1994), M67 (Mathieu et al. 1990), and NGC 1502 
(Mayer et al. 1994).

The majority of binary systems observed in open 
clusters are thought to be primordial, but
there is no preferred formation mechanism for multiple systems 
(dynamical or primordial) at present. Most of the multiple systems studied
are hierarchical because of the intrinsic stability of these systems. 
The origin of observed multiple systems has
not been clear since the beginning of the study of these systems. 
Duquennoy (1988) analyzed a sample of 17 systems (14 triples and 3 quadruples)
in the solar neighbourhood. He obtained a linear correlation
between the logarithm of the inner and outer binary period.
This was interpreted as an indication of preferential 
primordial origin for these systems. From a theoretical point of view,
Boss (1991) has suggested that the formation of hierarchical systems occurs
during the collapse of protostellar cores. Recently, Mermilliod et al. (1994)
have found significant period ratios ($X = P_{out}/P_{in} \simeq 250$)
in clusters which suggest a dynamical origin. Although the question
of the origin of these hierarchical systems is far from being answered,
we assume here
that all the hierarchical systems formed in open clusters have a purely 
dynamical origin. 

The formation and dynamical evolution of hierarchical systems in open
clusters can be studied
within the context of $N$-body simulations because complex interactions
between stars can readily be followed in detail by 
numerical methods. This approach has recently been adopted
(Aarseth 1996a, Kiseleva et al. 1996, Eggleton \& Kiseleva 1996, Kiseleva 1996).
In this paper the results of almost a hundred cluster models are analyzed
with the purpose of studying the formation, evolution and final destinies 
of hierarchical systems in clusters. These models have been 
obtained using direct $N$-body integration by the standard workstation code
NBODY5 (Aarseth 1985, 1994) and the more recent
NBODY4 (Aarseth 1996b).

NBODY5 consists of a fourth-order
predictor corrector scheme with individual time steps. 
In order to account for stellar evolution and mass loss (stellar winds and
supernova events), we use the fast fitting functions of
Eggleton, Fitchett and Tout (1989) and Tout (1990) 
for population I stars.

NBODY4 is based on 
the so-called Hermite scheme (Makino 1991) which forms the basis of a
new generation of special-purpose computers (Makino, Kokubo \& Taiji 1993).
Mass loss by stellar winds is now treated according to a modified
Reimers (1975) expression (Tout 1990).
Chaotic tidal motion (Mardling 1995),
tidal circularization (Mardling \& Aarseth 1996), 
exchange of mass (Roche overflow) in binaries 
and magnetic braking are also included. 

The models with NBODY5 have been studied on a DEC 2100 4/275 AXP system. 
All the calculations performed with NBODY4 were made at Cambridge
on the HARP-2 computer.

\section{Stability of hierarchical systems}
The main aim of this paper is to study the formation and 
evolution of hierarchical systems in open clusters. Considered as
isolated from field stars, such systems can survive over a 
long time. Small non-hierarchical systems, except in a few special cases,
are always unstable in the long term (Marchal 1990). Even for hierarchical 
triple systems, stability is not an easy question. There are a number of 
criteria to identify triple systems as stable or unstable, obtained 
analytically or numerically (see e.g. Kiseleva 1996).
For historical reasons, the models described here employ two criteria
which are in fact fairly similar for most mass ratios involving stars.
Thus the older NBODY5 code employs the Harrington (1975) criterion in
the form

\begin{equation}
F^{min} = A \left (1 + B\; log 
\left ( \frac{1 + m_3/(m_1 + m_2)}{3/2} \right )\right ),
\end{equation}
with $A = (2.65 + e) (1 + m_3/(m_1 + m_2))^{1/3}$ modified by the inner
eccentricity, $e$, according to Bailyn (1984) and $B = 0.7$.
Here $F^{min}$ denotes the critical ratio of the outer periastron
distance of the mass $m_3$ to the inner apastron distance of $m_1 + m_2$.

In the NBODY4 code, we adopt the stability criterion of 
Eggleton and Kiseleva (1995, hereafter EK).
In either case, we use
their definition of stability: that a hierarchical triple system is stable 
if it persists continuously in the same configuration (which 
excludes exchange and disintegration).

The stability criterion of EK used
for the NBODY4 models is based on the critical period ratio for stability,
$X^{min}_{0}=P_{out}/P_{in}$. It was derived from
a numerical study which examined
a wide range of parameters: (a) eccentricities of both
inner and outer orbits, $e_{in}$ and $e_{out}$; (b) relative inclinations
of inner and outer orbits, from prograde to retrograde;
(c) initial relative phase; (d) both mass 
ratios, $q_{in}$ = $m_{1}/m_{2}\:\geq\:1$ and 
$q_{out}=(m_{1}+m_{2})/m_{3}$.
This criterion can be written in two forms. Let $Y_{0}^{min}$ again be 
the critical ratio $F^{min}$.
For computational purposes $Y_{0}^{min}$
is a more relevant parameter, since it depends {\it only}
(to a certain level of approximation) on the two mass
ratios. The period ratio $X_{0}^{min}$ also depends on the two eccentricities
and is more useful for observed triples, since their periods and
eccentricities, rather than semi-major axes, are normally determined by
observation. Note that the inclination and phase are not included
explicitly in the criterion.
The two forms of the stability criterion are given by:

\begin{equation}
(X_{0}^{min})^{2/3}\:=\:(\frac{q_{out}}{1\;+\;q_{out}})^{1/3}\:
                      \frac{1\;+\;e_{in}}{1\;-\;e_{out}}\:Y^{min}_{0} \,,
                      \label{f1}
\end{equation}
\begin{equation}
Y^{min}_{0}\:\approx\:1\;+\;\frac{3.7}{q_{out}^{1/3}}\;-
                      \;\frac{2.2}{1\;+\;q_{out}^{1/3}}\;+
                      \;\frac{1.4}{q_{in}^{1/3}}\:\frac{q_{out}^{1/3}\;-\;1}
                      {q_{out}^{1/3}\;+\;1} \,.  \label{f2}
\end{equation}

This criterion discriminates between those 
triples that are likely to last a long time in the absence of strong external
perturbations and those which may break up rapidly.
Comparison of the two criteria used shows reasonable agreement, except
for the less probable case of a massive outer component.

There are two main conditions for identifying the `birth' of a
hierarchical subsystem in our simulations:
\begin{description}
\item[$\bullet$] The inner binary must be hard and the
                 outer component must also form a hard binary with the
                 inner binary.
\item[$\bullet$] The criterion (\ref{f2}) for hierarchical stability must be
                 satisfied.
\end{description}

If both the above conditions are satisfied simultaneously the subsystem is
considered as a hierarchical triple (or higher multiplicity) system and a 
number of parameters of this system are recorded. The subsequent treatment
consists of combining the inner binary into one body to permit a 
two-body treatment (recursively if double hierarchy). 

A hierarchical system is terminated when one of the
following situations occurs:
\begin{description}
\item[$\bullet$] The stability criterion (\ref{f2}) is violated because, 
                 for example, of significant changes of $e_{out}$ due 
                 to small secular external perturbations.
\item[$\bullet$] The external perturbation
                 exceeds a critical value.
\item[$\bullet$] Effects of stellar evolution become important or the
                 predicted pericentre distance becomes too small.
\end{description}
In the last case, however, the same hierarchical system usually appears again, 
with slightly different parameters (e.g. during gentle mass loss) and we 
take this situation into account in the statistical analysis of results.

In the present calculations it is possible to produce triple systems, quadruple
systems and double hierarchies. Triple systems consist of an inner binary and
an outer single star, quadruple systems also have a binary as an outer body and
in double hierarchies we have a nested system (binary + outer body + another
outer body) with up to six stars. Hence, we can
study even hierarchical sextuple systems.
Recently Mardling and Aarseth (1996) have developed
a new stability criterion based on a chaos description (Mardling 1995) which
contains the inner and outer eccentricities explicitly as well as the mass ratio
$m_3/(m_1 + m_2)$ for coplanar orbits.
Except for large $e_{out}$, it gives values that are somewhat smaller than EK.

\section{Cluster models}
In order to obtain
realistic results it is important to choose a general mass
function (hereafter IMF). In all the models the IMF used is described by:
\begin{equation}
f(m)\:=\:\frac{0.3\;{\rm X}}{(1\:-\:{\rm X})^{0.55}} \,,
\end{equation} 
where $X$ is a random variable in $[0, 1]$. This IMF is a fit
to Scalo's (1986) results. For binaries we introduce a correlation 
$(m_{1}/m_{2})' = (m_{1}/m_{2})^{0.4}$, subject to the sum being constant,
which yields mass ratios closer to unity than for random samples.

The initial NBODY5 models have a mass
range of 0.1-15.0 $m_{\odot}$ for the single stars.
Spherical symmetry and constant density are assumed, with the
ratio of total kinetic and potential energy fixed at 0.25. 
All these models have 
random and isotropic initial velocities. 
Stars outside twice the classical tidal radius are assumed to escape and
are removed from the calculations. 

The models with NBODY4 have a mass range of 0.2-10.0
$m_{\odot}$ for the single stars. Initial coordinates and velocities are 
generated from an equilibrium King model in an external galactic field 
(Heggie and Ramamani 1992). In each code
the cluster is assumed to be in a circular orbit in the solar
neighbourhood, with a linearized tidal force added to the equations of
motion (Aarseth 1985, 1994). 

The probability of formation of hierarchical systems
in a cluster with only single stars is very small; hence 
in order to produce a relatively large number of such systems, a
significant initial binary population is needed
(de la Fuente Marcos 1996). 
NBODY4 models can
have an arbitrary number of binaries; for example models with 5100 stars have
100 primordial binaries. However, for NBODY5 models ($N = 100, 500$ and 1000) 
two binary fractions,
10 \% and 50 \%, have been studied. Apart from the binary fraction some 
other parameters must be specified for the initial binary population; 
in particular,
the semi-major axis and eccentricity must be chosen. For an 
initial population of hard binaries the semi-major axis is about 
$- G M^{2} / 4 E N$, where $G$ is the gravitational constant, $M$ is the
total mass, $E$ is the energy.
The semi-major axis is taken from a uniform distribution: 
$a_{b} = a^{0}_{b}\:10^{-q}$, where $a^{0}_{b}$ is a parameter 
whose value is 1/$N$ in units of the virial radius, and $q$  is equal to 
$X\log{R}$. $X$ is a random number uniformly distributed in the interval 
$[0, 1]$ and $R$ is a parameter. The latter gives the 
spread in semi-major axes and thus the spread in 
energies and periods.
For example, in the $N = 5100$ models with NBODY4 the binary semi-major axis is
in the range 0.1 -- 100 AU. For NBODY5 models, 
all the runs have been repeated with
$R = 5, 10, 50, 100$,
with maximum values of the semi-major axes 
in the range 230 -- 1000 AU.
Moreover, all the runs 
have been computed twice: one time with the chain regularization excluded and
another time with this option included. Chain regularization 
(Mikkola \& Aarseth 1990, 1993) is a numerical 
treatment of close encounters in compact subsystems in which the external
perturbation due to nearby stars is taken into account.
 
\section{Results}
In this section we discuss some representative results.
We are mainly concerned with results which can be checked directly with 
observational ones. 
  
\subsection{Overall evolution}
Although there are three sets of different simulations, we find  
some common features. As expected, the majority of 
the systems formed are triple, but some NBODY4 models show
almost the same tendency to form quadruples. Hierarchies do not form
at a preferred cluster evolution stage. Usually in 
clusters with primordial binaries the first hierarchies appear at about
0.025 - 0.05 $T_{cl}$, where $T_{cl}$ is the total life-time of the cluster, 
and in the cluster remnant (when $N$ is a few tens) there are sometimes
long-lived hierarchies.
The distribution of inclinations for hierarchies is fairly symmetrical,
with a pronounced peak centred on $90^{\circ}$ (Kiseleva 1996).
In large clusters at a late stage of the evolution, many short-lived systems
form via repetitive triple-binary and triple-triple exchanges. 
Sometimes a hierarchical system leaves the cluster, escaping before 
the disruption of the system takes place.
Poor clusters ($N$ = 100) with wide binaries
do not show any tendency to form hierarchical systems
since we require the outer orbit to be a hard binary.
Sometimes the outer body is a collapsed
object (white dwarf).
The largest number of hierarchies
observed at the same time is five. 

First we present some results for NBODY5 models. 
In the triple systems which form, the inner binary is usually not primordial.
The quadruple systems found in poor clusters 
are more eccentric in models with no chain treatment. 
These systems form typically at the late stages
of cluster evolution in rich clusters and inside the cluster core. 
For systems which include the chain treatment the quadruple systems
formed in rich clusters are more eccentric. They also form preferentially
inside the core. 
For triple systems the outer star usually has a mass less than one-third
of the binary mass and sometimes 1/15, when the outer body is a 
collapsed object.  
The typical life-time for the systems formed is a 
few million years, or 10$^3$ outer periods. The inner binaries of 
the hierarchies have periods from a few days to 10$^{3}$ days. 

NBODY4 models also show preferential formation of triples but now 
quadruples are relatively abundant (up to 40 \% of all hierarchical
systems in $N = 5100$ models).
Quadruples tend to form at late stages of the cluster evolution.
On the other hand, the corresponding percentage of systems with primordial
inner binaries is significantly greater than for the NBODY5 models 
(about 70 \% vs 40 \%) because of the shorter periods.
Recent models include a new technical feature of so-called double
hierarchies.
In such configurations an outer body (single or binary) is added
to an existing stable triple or quadruple, provided the stability criterion
is satisfied.
These systems tend to occur during very late stages when the
core has expanded significantly; however, they tend to be short lived in
the models studied so far with termination due to significant external
perturbation.

The formation of hierarchical systems may have interesting observational
consequences (Aarseth 1992). In such
models, after the formation of a hierarchical triple system the
pericentre of the outer body is usually close enough to the inner binary
to perturb its eccentricity. This systematic perturbation often promotes
a stellar collision of the inner binary. This process has been observed 
in several NBODY5 models with $N$ = 1000 and 50 \% primordial binaries,
and is quite common in NBODY4 models which contain binaries with periods
of days.

\subsection{The Period -- Eccentricity plane}
One of the most interesting diagrams for displaying the results of
binary observations is the plot of the eccentricity versus the logarithm
of the period because
it provides more insight
into binary astrophysics than other distributions of orbital elements 
and is expected to be equally useful in the study of hierarchies.
We have plotted the eccentricity of the outer body 
(single or binary) against the logarithm of its period (in days).
The results depend on the type of model
and also on the membership and initial 
orbital elements of the binary population.
As remarked above, our models can be classified into three different 
groups but several features are common to all:
\begin{description}
\item[$\bullet$] A lower cut-off period below which no orbits are observed. 
\item[$\bullet$] An upper cut-off period above which no orbits are 
                 observed. 
\item[$\bullet$] A zone of forbidden eccentricities
                 for systems in which the inner binary is not primordial.
\end{description}
%
%---------------------------------------------------------------------------
\begin{figure}[hbt]
%  \vspace{5cm}
  \centerline{\hbox{
  \psfig{file=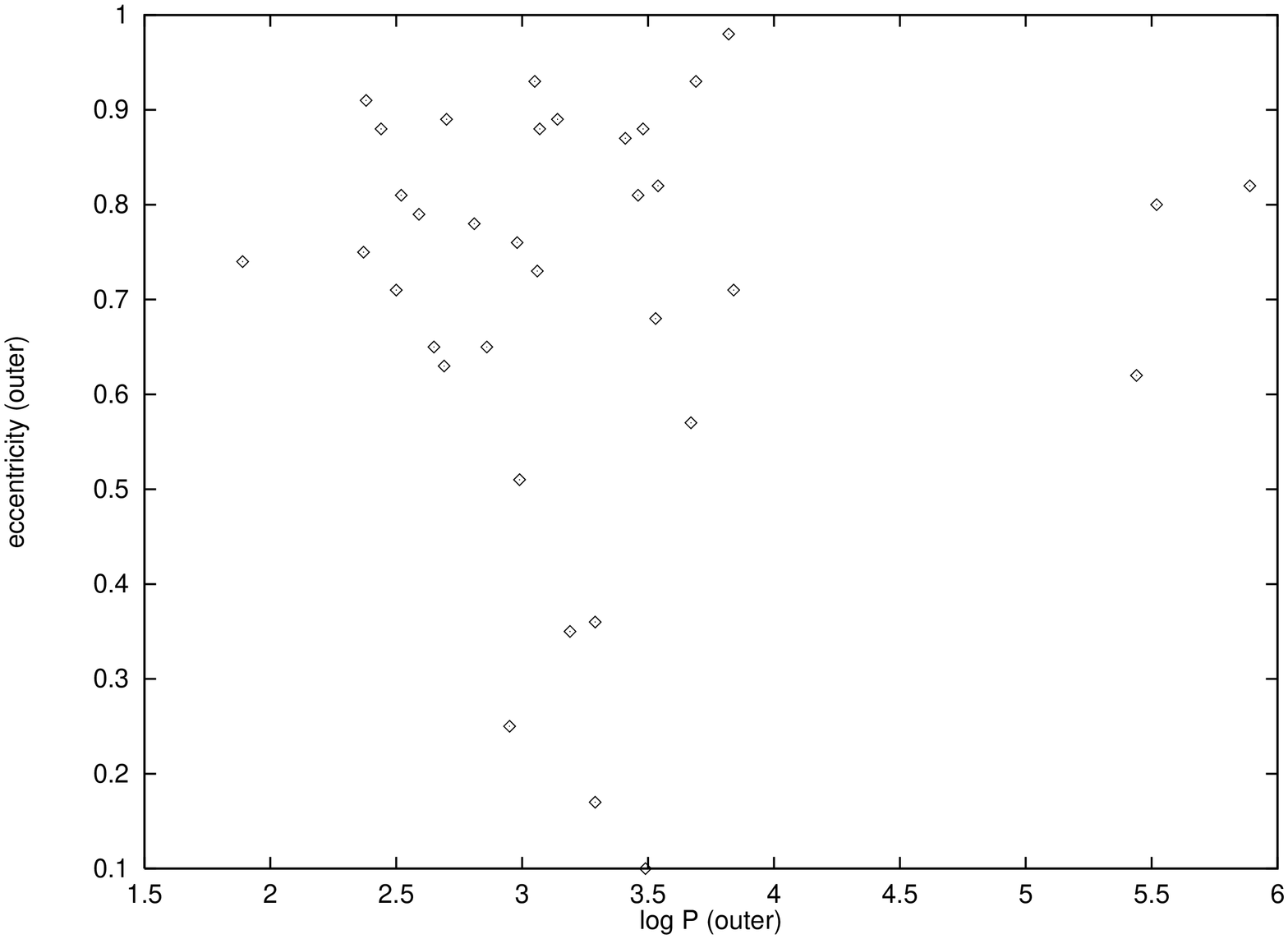,height=6cm,width=6cm,angle=-90}
  \psfig{file=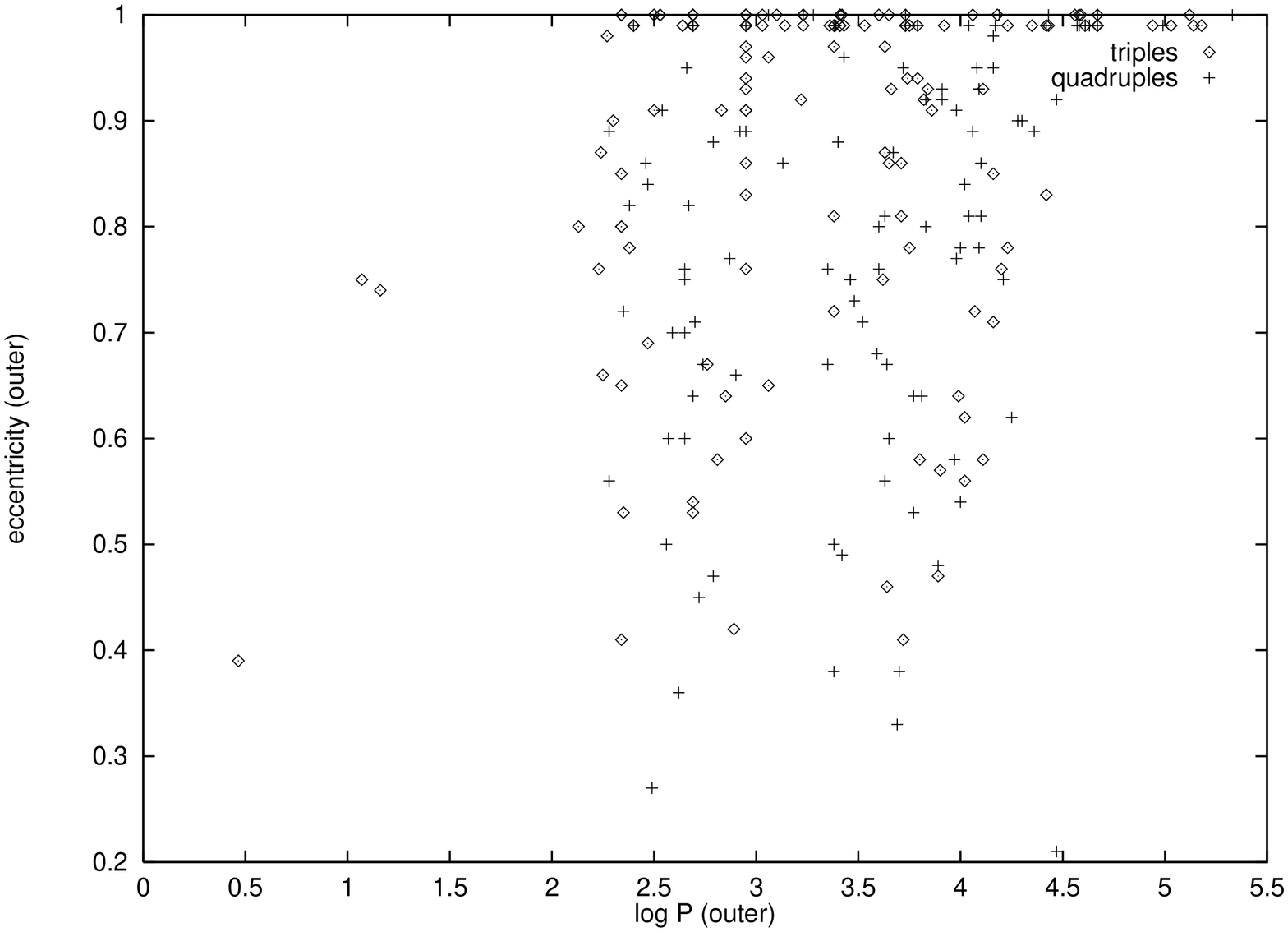,height=6cm,width=6cm,angle=-90}
  }}
  \caption{Eccentricity -- period diagram
           of the outer binary in days.
           Left panel is for an NBODY5 model, the right one is for
           an NBODY4 model. High eccentricity systems are 
           short-lived in all the models.}\label{ep}
\end{figure}
%---------------------------------------------------------------------------
%

{\it The short period cut-off.} 
Its value is dependent on the initial values
of the initial primordial binary parameters. For 
the NBODY4 models common envelope evolution is
included so a few systems are observed
below the cut-off period; most of these
suffer a physical collision which sometimes produces a blue straggler.
For models without Roche overflow
and common envelope evolution wide inner binaries 
implies longer outer periods; i.e. models with very hard binaries 
form hierarchical systems which permit shorter periods for the outer body.
Moreover, the cut-off is larger if the inner binary
is not primordial. In fact, even for the NBODY4 models, we observe
a clear cut-off in the period ratio with no exceptions, due to common envelope
evolution or collisions.
Another interesting feature is the dependence on the nature 
of the outer 
body (single or binary). The cut-off is larger for systems in which the
outer body is a single star.

{\it The long period cut-off.}
This is mainly due to the stability limit against
perturbations from the neighbouring stars. Models with wider binaries show
higher values of the upper period cut-off. The width of the period
distribution depends strongly on the initial distribution of the 
semi-major axes of the binaries; small binaries generate sharp distributions
of periods for the outer body. The high cut-off also depends on the
nature of the inner binary. Systems with an inner primordial binary 
show slightly higher periods and the same is found for systems in which the
outer body is also a binary.

{\it The forbidden region.}
A dependence of the eccentricity distribution on the character of the
inner binary is observed. When the inner binary is not primordial no
systems are observed with eccentricities smaller than 0.20. Compared with
real systems, this can be considered as a limit below which all the systems
observed should contain a primordial binary. This must reflect the fact 
that non-primordial binaries, although usually formed by exchanges or
disruptions of primordial systems, are wider and less energetic than 
primordial ones. The existence of real systems below this limit with a 
non-primordial
inner binary may be explained as an effect of tidal interaction.
There is also a correlation between the membership of the
cluster and the lower eccentricity observed. Finally, systems formed in poor 
clusters are more eccentric on average. 

\subsection{The (log P$_{b}$, log P$_{o}$) plane}
\begin{figure}[hbt]
%  \vspace{5cm}
  \centerline{\hbox{
  \psfig{file=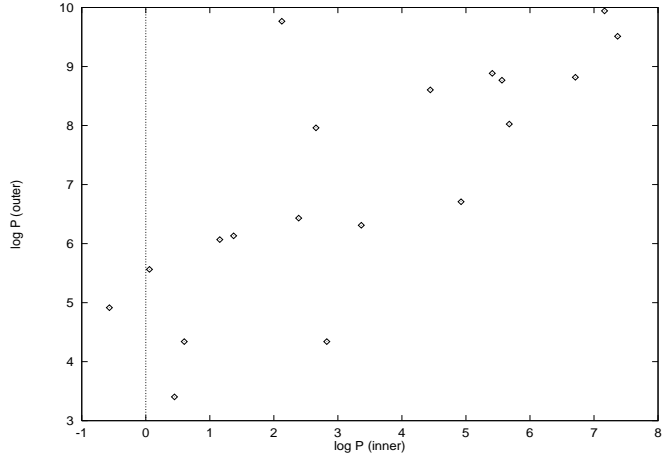,height=6cm,width=9.0cm,angle=-90}
  }}
  \caption{Period diagram for the observational sample of multiple 
           systems listed in Duquennoy (1988). Note 
           the larger outer periods in comparison with fig. \ref{pp} because 
           most of them are estimated parameters in the original paper.}
           \label{ppobser}
\end{figure}

In order to compare our results with those from the 
observational literature, we consider 
the plane log P$_{O}$ vs log P$_{B}$,
where P$_{O}$ is the outer orbital period and P$_{B}$ the inner period. 
We compare mainly with a sample of triples (fig. \ref{ppobser}) in the solar 
neighbourhood by Duquennoy (1988). He finds a linear correlation
for a sample of 13 triples with components of solar type. The slope
of the straight line is 0.68 with a correlation coefficient of 0.91. 
From the relation between the periods he concludes that triple systems are
rarely formed by capture, but rather by fragmentation processes, 
because capture would produce random combinations of periods. He suggests
dynamical evolution as another possible cause in order to 
reproduce the observed distribution of periods. However, triples in our models
have a purely dynamical origin and we find a similar behaviour.
%
%---------------------------------------------------------------------------
%-----------------------------------------------------------------------------
%

\begin{figure}[hbt]
%  \vspace{5cm}
  \centerline{\hbox{
  \psfig{file=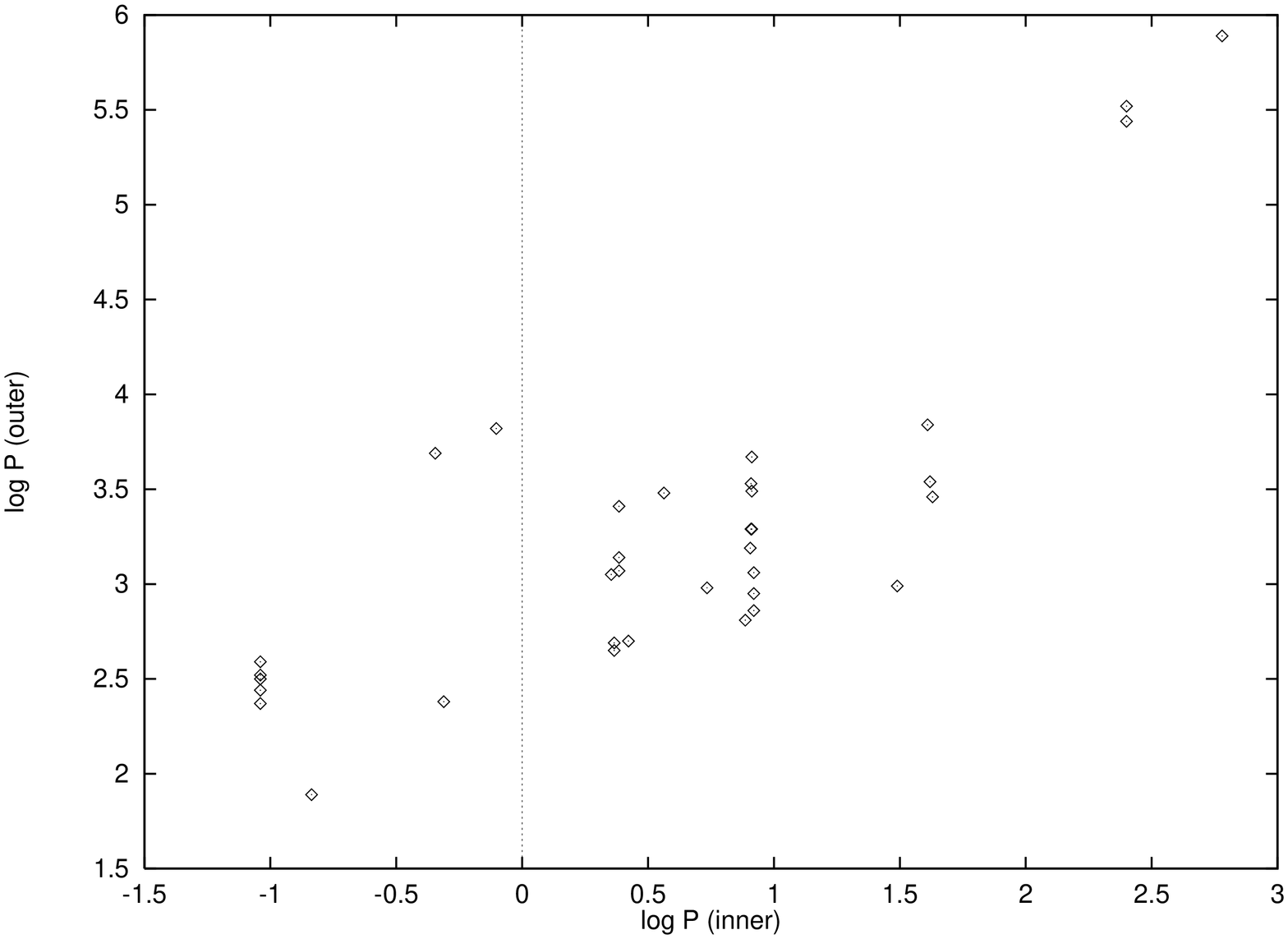,height=6cm,width=6cm,angle=-90}
  \psfig{file=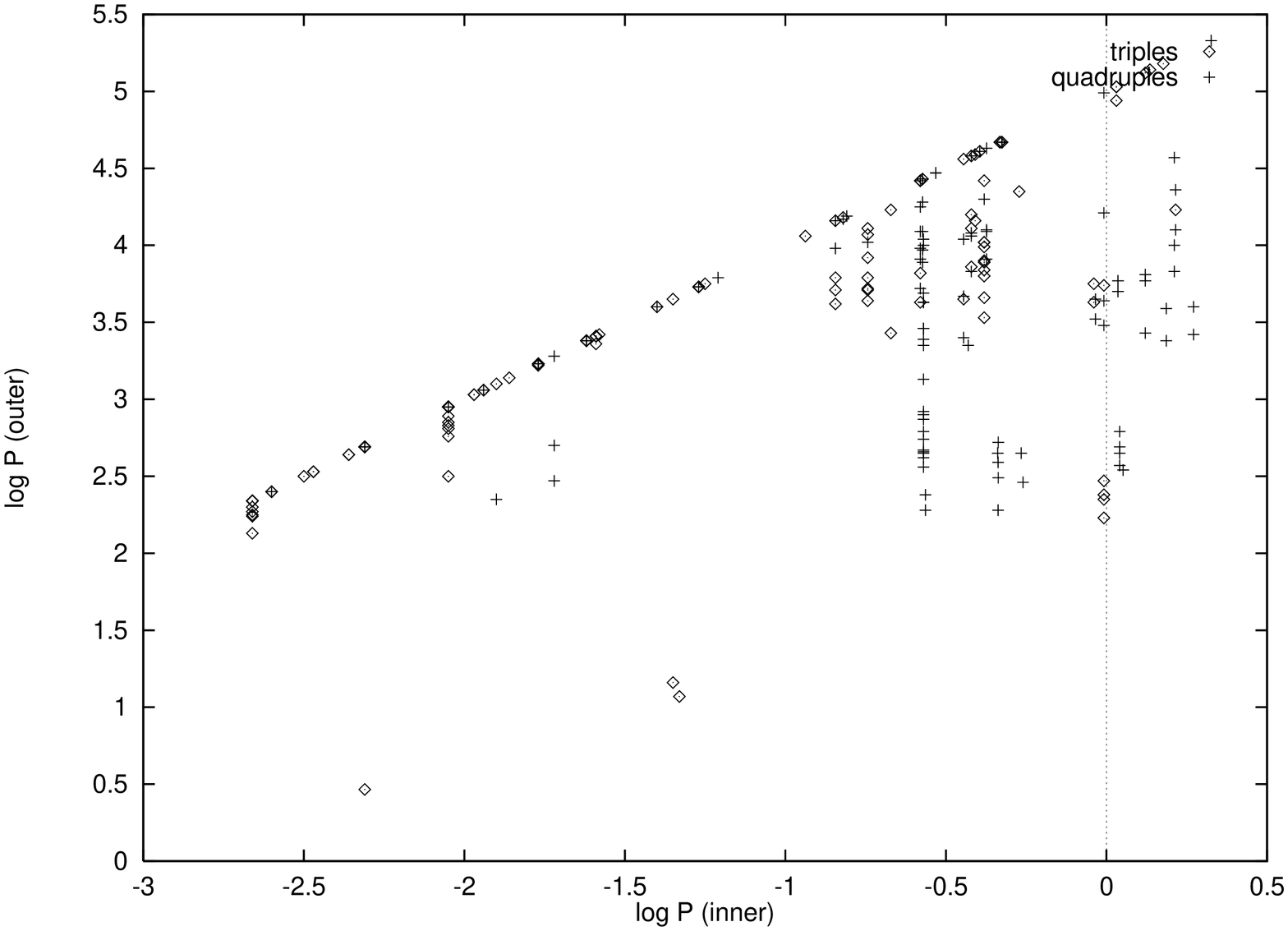,height=6cm,width=6cm,angle=-90}
  }}
  \caption{The left figure shows the plane log P$_{O}$ vs log P$_{B}$ for
           a model studied with NBODY5.
           The right figure is for a model
           performed with NBODY4. The linear upper contour is 
           due to the criterion for stability. Note that inner binaries in
           NBODY4 have smaller periods due to the astrophysical 
           phenomena included (see the text).}\label{pp}
\end{figure}

For an NBODY5 model with 1500 stars
and 500 primordial binaries (fig. \ref{pp}) 
we obtain a slope of 0.68 $\pm$ 0.09. 
For an NBODY4 model with 5100 stars and 100 primordial 
binaries we obtain a slope of 0.53 $\pm$ 0.03.
Both results include
systems in which the outer body is either a single or a binary. However,
the number of quadruple systems in the first case is about 12 \% (4:34) and
about 45 \% in the second. We analyze the 
two subsamples (triples and quadruples) in the latter model. For triple systems
we obtain a slope of 0.63574 $\pm$ 0.00008 and
for quadruple systems we find 0.4219 $\pm$ 0.0001.  
The results suggest that there are two kinds of correlations depending
on the nature of the outer body.

Figure \ref{pp} shows
the plot for an NBODY4 model; the upper contour is clearly linear and is
connected with the stability criterion. Our results seem to be 
compatible with those from the observations in spite of our theoretical 
sample of hierarchical systems containing a wide range of masses and different
stages of evolution.
The lack of systems with large outer periods but small inner period is of
theoretical and observational interest.
This is probably connected with the formation
mechanism in which other stars act as perturbers during wide encounters.
Thus it has been noted from the calculations that 
hierarchical formation mainly takes place when two or more binaries suffer
a close encounter.
%
%---------------------------------------------------------------------------
%---------------------------------------------------------------------------
%
It would be desirable to compare our results with those from systems 
observed in real open clusters. 
Unfortunately, there are only three such systems with orbital
elements for both the inner binary and the outer body. Two are
in the Hyades, vB 75 and $\mu$ Orionis; the latter is a quadruple.
The other system is a quadruple in NGC 1502, SZ Cam.

\subsection{The (log P$_{o}$, M$_{b}$/M$_{o}$) plane}
Although this plane (fig. 4) also contains much astrophysical information,
it is not easy to obtain the relevant data from observations. Here M$_{B}$ 
is the mass of the inner binary and M$_{O}$ is the mass of the outer body.
For the NBODY4 models the majority of the systems formed fall in the
mass ratio range 1--2.2. There are a few systems with higher ratios
and the upper cut-off is about 14. The number of systems with mass ratios
below 1 is very small. As for the other correlations 
there is a clear distinction
between quadruple systems and triples. The range of mass ratios for 
quadruples is 0.4--2.2, which is expected because in most cases both
binaries are primordial. For triple systems there are several systems 
with ratios greater than 2.2 but none below 0.8. As regard the 
nature of the inner binary, systems with no primordial binaries have 
a range 0.8--3.0 with only one system above (14). For systems 
with a primordial inner binary the range is wider.
%
%---------------------------------------------------------------------------
\begin{figure}[hbt]
%  \vspace{5cm}
  \centerline{\hbox{
  \psfig{file=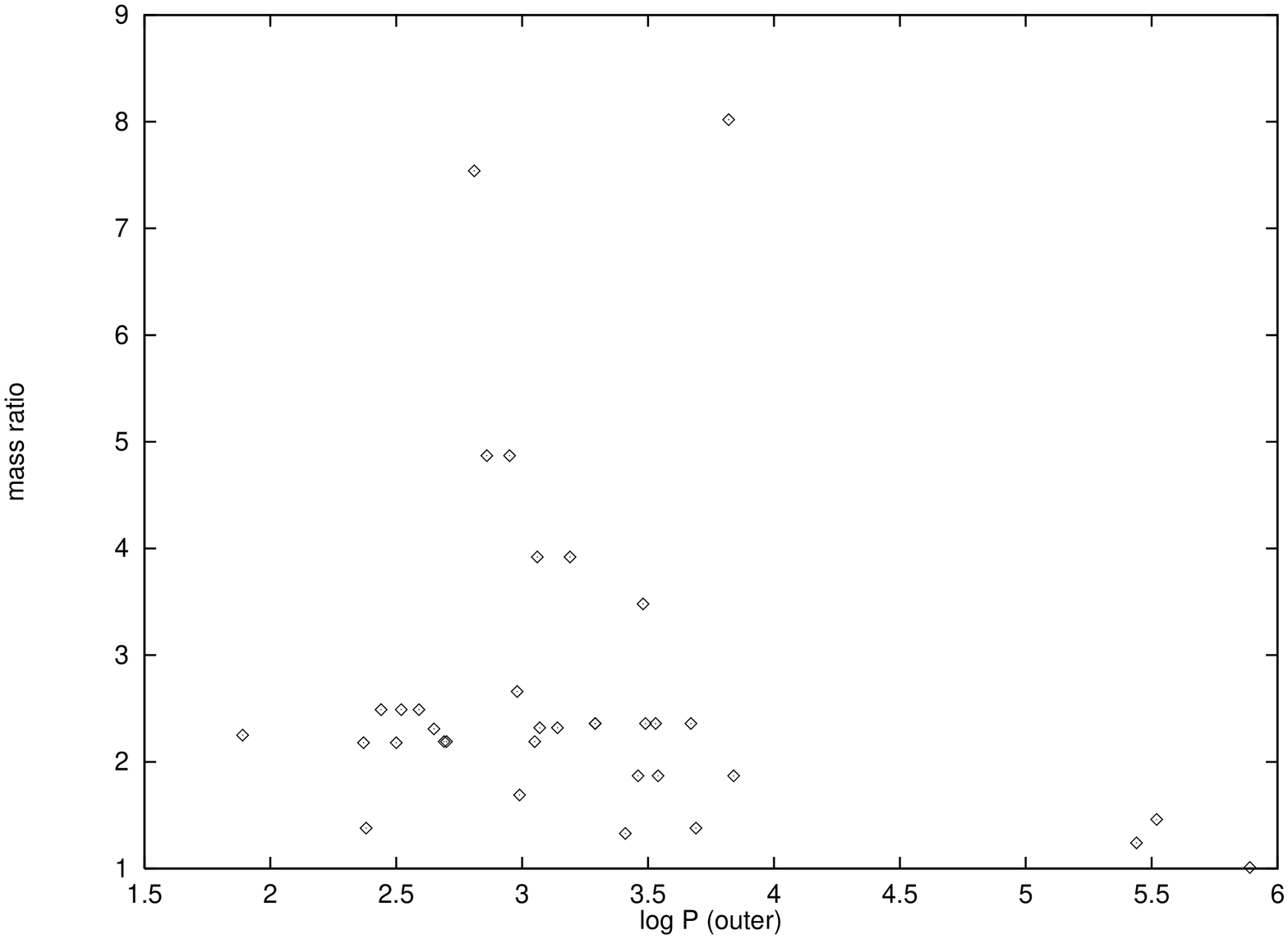,height=6cm,width=6cm,angle=-90}
  \psfig{file=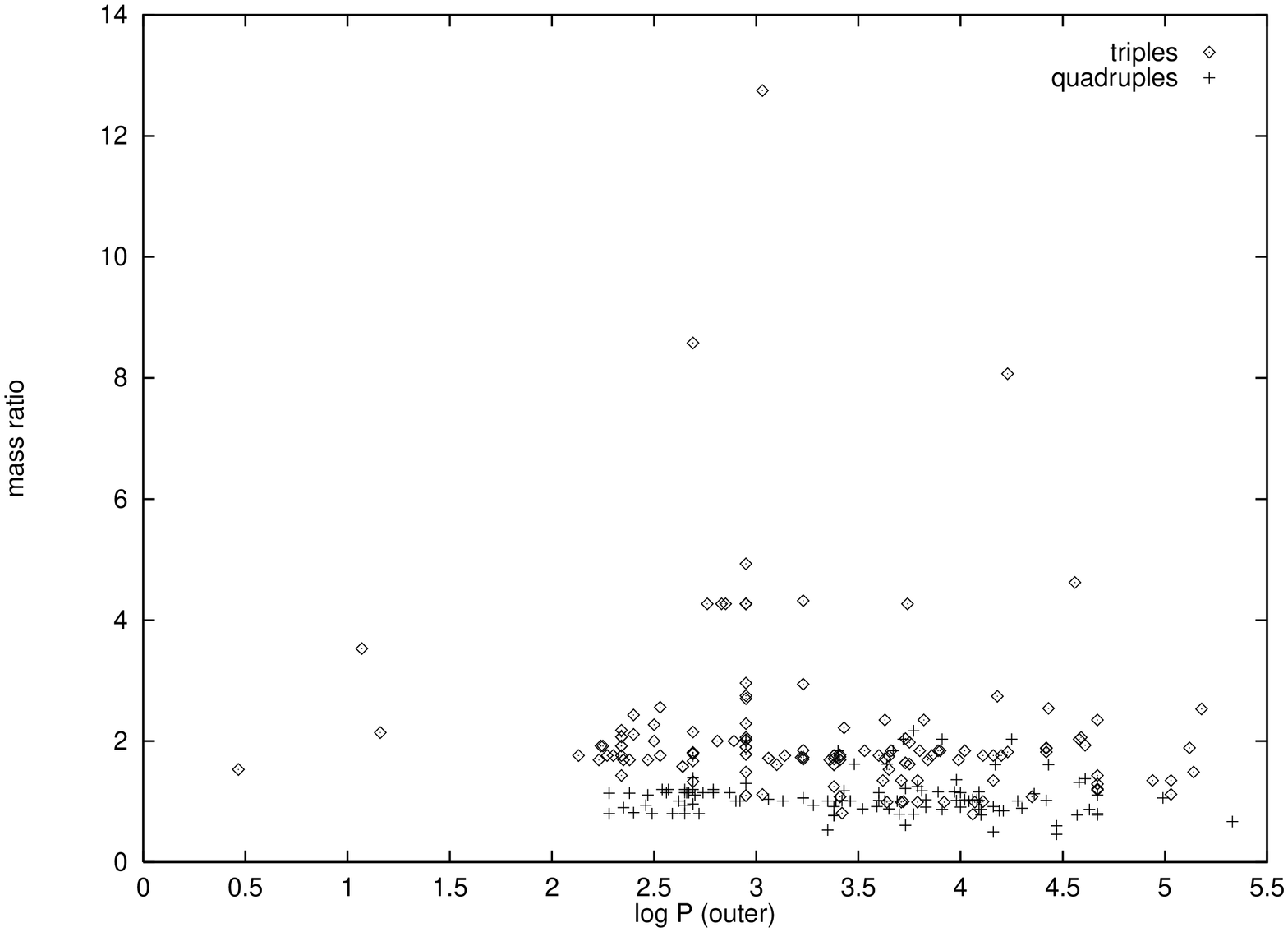,height=6cm,width=6cm,angle=-90}
  }}
  \caption{Mass ratio versus logarithm of the period of the outer body
           for an NBODY5 model (left) and an NBODY4 model.
           Note the sharp mass ratio for quadruple systems in the 
           latter.}\label{mp}
\end{figure}
%---------------------------------------------------------------------------
%

\section{Tidal dissipation in triple systems}
Neither primordial binaries, nor those that are formed in clusters
later (for example by exchange), have exactly circular
orbits to start with. However, tidal friction will circularise
binary orbits on a rather short time-scale compared with the nuclear time-scale,
provided that at least one star of the binary has a radius comparable
to the separation between binary components and  the dynamical
influence of other stars on the binary orbit is negligible. The
situation can be very different for binaries within hierarchical
triple systems, especially those where the inner and outer orbits are
nearly perpendicular (as our  numerical simulations
shows, not a rare situation).
It can be shown  analytically and numerically (Marchal
1990, Heggie 1996, Kiseleva 1996) that for non-coplanar 
triple systems there is a quasi-periodic change of the inner eccentricity 
(on a time-scale $\sim P^2_{out}/P_{in}$) during which it reaches a maximum 
value $e_{\rm in}^{\rm max}$.  This value only depends on the inclination $i$ 
between the two orbital planes; other parameters affect only the time-scale.
If $i \approx 90^o$, $e_{\rm in}^{\rm max} \approx 1$ and the two stars may 
collide or suffer a strong tidal interaction. This
effect cannot be neglected in numerical studies of triple stars in
clusters. The combined influence of tidal friction and of the third
component on the  binary orbit may  produce  interesting and
even dramatic results, such as for example a severe shrinking of the
orbit.  

In order to investigate  
 the interaction between tidal friction and the gravitational
dynamics of point masses in a hierarchical triple system we consider  
two well-known isolated triples $\beta$ Per (Algol) and $\lambda$ Tau,
which have well-defined orbital parameters. 
 The influence of the distant third body induces eccentricity in the orbit of
the close pair.
Even if the third body is distant, as in $\beta$ Per ($P_{\rm out}
/P_{\rm in} \approx 237$), its effect need not be small. Because
 the observed  $i \approx 100^{\circ}$ (Lestrade et al. 1993), in the absence
of a dissipative process like tidal friction the eccentricity of the inner
pair should cycle between 0 and $\sim 1$, on a time-scale of $\sim 10^3$
 years (fig 5, left panel). 
Since the observed eccentricity is $\sim 0$, and presumably
has been small throughout its current phase of Roche overflow, 
 tidal friction must act successfully in this system. The right panel
 of fig. 5 shows how tidal friction (included in the equations of motion
 of the three-body problem  as described by Eggleton 1996) can reduce the
 effect of the distant companion on the inner eccentricity. 
\begin{figure}[hbt]
% \vspace{5cm}
  \centerline{\protect\hbox{
  \psfig{file=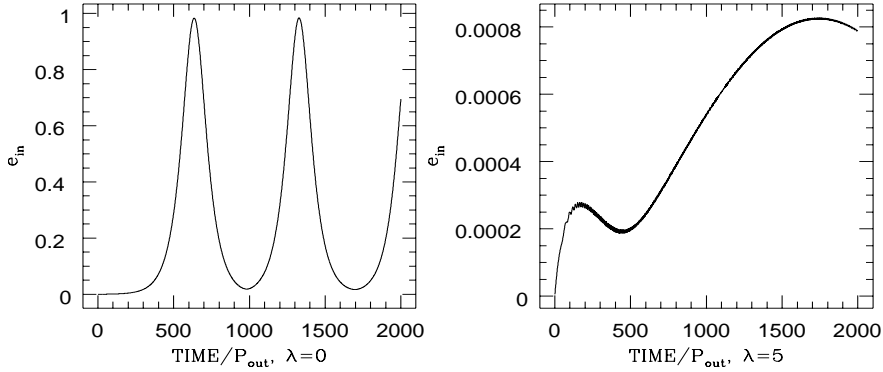,height=5cm,width=12.0cm,bbllx=25bp,bblly=430bp,bburx=575bp,bbury=690bp,clip=}
  }}
\caption{The effect of tidal friction on long-term modulations of the
 inner eccentricity in $\beta$ Per triple systems. The parameter $\lambda$
defines the strength of tidal friction and is described in Eggleton
(1996).}
\end{figure}
 The recipe for tidal friction 
dissipates orbital energy but conserves angular momentum; it
decreases the semi-major axis and orbital period of the inner binary, 
and hence increases the period ratio.
 In many cases this effect can be rather insignificant. 
 For  $\beta$ Per (and other similar systems with
 high inclination) we find a rather narrow range of
 tidal friction values which is not strong enough to prevent
 significant quasi-periodic variations of the inner eccentricity, and
 yet is strong enough to decrease sharply the binary semi-major axis every
 time the eccentricity reaches its local maximum. Figure 6 shows this
 effect. 
This is a  possible mechanism for the production of close
 binaries and/or other exotic objects, particularly in clusters. 
\begin{figure}[hbt]
% \vspace{5cm}
  \centerline{\hbox{
  \psfig{file=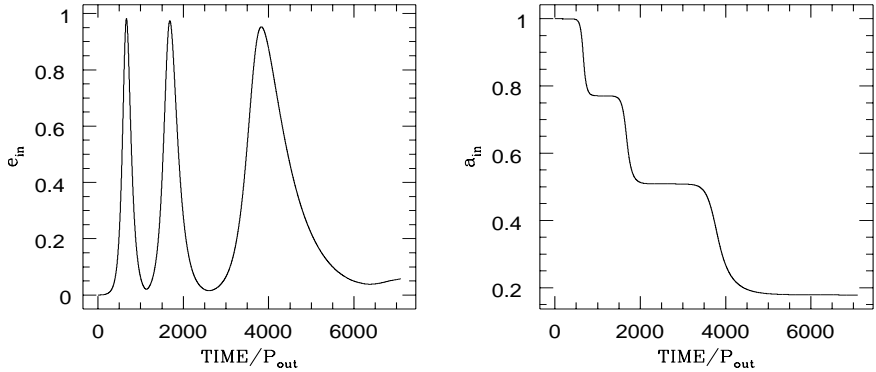,height=5cm,width=12.0cm,bbllx=25bp,bblly=430bp,bburx=575bp,bbury=690bp,clip=}
  }}
\caption{The possible shrinking of the binary orbit in a triple system
 like $\beta$ Per under the influence of rather weak tidal friction.} 
\end{figure}

\section{Conclusions}
This work has allowed us to discuss various aspects of
hierarchical systems in open clusters.
We have found that it is, indeed, possible to reproduce some observational
properties (such as the linear correlation of periods) of hierarchical systems 
as well as to predict some characteristics of these systems for observations.
Although the results described in
this paper are encouraging, it is still not clear how the fraction of
primordial 
binaries influences the formation rate of hierarchical systems and how the 
tidal effects, which were only discussed briefly here, affect the
stability of systems with short periods.
These questions will be left for future developments.

\section{Acknowledgements}
R.F.M. thanks the Department of Astrophysics of Universidad Complutense de
Madrid for providing excellent computing facilities. L.G.K. thanks
the NATO for a Collaborative Research Grant.
Our thanks are due to J.-C. Mermilliod who provided
most of the observational data
both personally and across his database (BDA) at Lausanne University.

\end{document}